\documentclass[%
 article,
 pra,
 amsmath,amssymb,
 aps,
 twocolumn,
]{revtex4-2}
\usepackage{placeins}
\usepackage{upgreek}
\usepackage{graphicx}
\usepackage{dcolumn}
\usepackage{bm}

\begin{document}

\bibliographystyle{rev_tex_x27.bst}


\title{A Quantum Optical Microphone in the Audio Band}

\author{Raphael Nold$^1$$^,$$ ^2$}
\author{Charles Babin$^1$$^,$$ ^2$}
\author{Joel Schmidt$^1 $$^,$$^2$}
\author{Tobias Linkewitz$^1 $$^,$$^2$}
\author{Mar\'ia T. P\'erez Zaballos$^3$}
\author{Rainer St\"ohr$^1 $$^,$$^2$}
\author{Roman Kolesov$^1 $$^,$$^2$}
\author{Vadim Vorobyov$^1 $$^,$$^2$}
\author{Daniil M. Lukin$^4$}
\author{R\"udiger Boppert$^5$}
\author{Stefanie Barz$^2$$^,$$^6$}
\author{Jelena Vu\v ckovi\'c $^4$}
\author{Christof M. Gebhardt$^2$$^,$$^7$}%
\author{Florian Kaiser$^1$$^,$$^2$}\email{f.kaiser@pi3.uni-stuttgart.de}
\author{J\"org Wrachtrup$^1$$^,$$^2$}%
\affiliation{$^1$3rd Institute of Physics, and Research Centre SCoPE, University of Stuttgart, Stuttgart, Germany}
\affiliation{$^2$Center for Integrated Quantum Science and Technology (IQST), Germany}%
\affiliation{$^3$The Old Schools, Trinity Ln, Cambridge CB2 1TN, Reino Unido, UK}
\affiliation{$^4$Ginzton Laboratory, Stanford University, Stanford, CA, USA}
\affiliation{$^5$Department of Pediatric Audiology and Neurotology, Olgahospital, Stuttgart, Germany}%
\affiliation{$^6$Institute for Functional Matter and Quantum Technologies, University of Stuttgart, Stuttgart, Germany}%
\affiliation{$^7$Institute of Biophysics, Ulm University, Ulm, Germany}%

\date{\today}

\begin{abstract}
The ability to perform high-precision optical measurements is paramount to science and engineering. Laser interferometry enables interaction-free sensing with a precision ultimately limited by shot noise. Quantum optical sensors can surpass this limit, but single- or multi-photon schemes are challenged by low experimental sampling rates, while squeezed-light approaches require complex optical setups and sophisticated time gating. Here, we introduce a simple method that infers optical phase shifts through standard intensity measurements while still maintaining the quantum advantage in the measurement precision. Capitalising on the robustness and high sampling rates of our device, we implement a quantum optical microphone in the audio band. Its performance is benchmarked against a classical laser microphone in a standardised medically-approved speech recognition test on 45 subjects. We find that quantum-recorded words improve the speech recognition threshold by $-0.57\, \text{dB}_{\text{SPL}}$, thus making the quantum advantage audible. Not only do these results open the door towards applications in quantum nonlinear interferometry, but they also show that quantum phenomena can be experienced by humans.  
\end{abstract}

\maketitle


Optical interferometry represents the gold-standard for measuring small displacements \cite{Coddington2009c}, refractive indices \cite{Cimini2019e,Kim2021a}, and surface properties \cite{Masters2020a,Conroy2008c}. A textbook example is the Michelson interferometer. Here, the best achievable phase sensitivity $\Delta\Phi$ with $N$ uncorrelated photons is given by the shot noise limit (SNL)
 \cite{Giovannetti2006b,Demkowicz-Dobrzanski2015b,Slussarenko2017c}, $\Delta\Phi =1/\sqrt{N}$.
Quantum optical sensors surpass the SNL by exploiting correlations and entanglement \cite{Demkowicz-Dobrzanski2015b,Slussarenko2017c,Caves1981a,Pan2012a,Chekhova2016c,Brida2010a,Ulanov2016a}. Theoretically, $N$-photon entangled quantum states can reach Heisenberg scaling, $\Delta\Phi =1/N$, which is the ultimate sensitivity limit \cite{Giovannetti2006b,Slussarenko2017c,Caves1981a}. This can be understood by considering the $N$-photon entangled state as a single entity in which all photons accumulate a collective phase \cite{Dowling2008b}, $N\cdot \Delta\Phi$. Importantly, information about each $N$-photon quantum state is extracted in a single experimental repetition. This is different from the classical case based on $N$ uncorrelated photons that acquire phase shifts individually, thus necessitating $N$ single-photon measurements. The fact that a measurement's noise power increases with the square root of the repetitions, makes it clear that the quantum strategy can improve the signal-to-noise ratio (SNR) by a factor of $\sqrt{N}$.

Although quantum optical phase sensing has been implemented with multi-photon path entangled states \cite{Cimini2019e,Afek2,Ono2013c}, the achievable low repetition rates hinder the development of applications. This is mainly a conceptual issue. Traditional schemes require measuring whether an even or odd number of photons left each interferometer output port \cite{Birrittella2021}. For $N=2$, this imposes the use of single-photon detectors (SPD) to distinguish between zero and one photons, while experiments with $N>2$ necessitate photon number resolving detectors (PNRD) \cite{Malik2016a,Yao2012}.  The associated measurement rates tend to be slow as these detectors show low optical saturation levels \cite{Hadfield2009} and complex coincidence counting is involved \cite{Slussarenko2017c}.

The saturation problem can be overcome using standard photo detectors in pulsed squeezed light experiments, however due to the pulsed regime, the effective measurement duty cycle is typically very low, thus limiting overall sampling rates \cite{Kim1994a,Eckstein2011a,Chatzidimitriou-Dreismann1987a}. 100\% duty cycle is obtained with continuous-wave squeezed light, however it comes at the cost of complex setups \cite{Boyer2008b,Vahlbruch2016a} and instabilities at high pump powers \cite{Moerner1994a,Kaiser2016b,Kashiwazaki2020a}. In this field a challenge remains the sensitivity of squeezed states of light to loss \cite{Andersen2016b,Schnabel2017b,Wolfgramm2010a,Demkowicz-Dobrzanski2012b}, which compromises the overall sensitivity.

Recent approaches rely thus on nonlinear interferometry in SU(1,1) interferometer arrangements \cite{Chekhova2016c,Hudelist2014a}, but these schemes face challenges in addressing common mode noise (pump laser power fluctuations) \cite{Chekhova2016c,Hudelist2014a,Szigeti2017a}.

Alternative measurement concepts have been inspired by fundamental quantum optics research in the 1990s, where the concept of ``induced coherence without induced emission'' was introduced as an elegant way to create photonic superposition states \cite{Zou2, Wang} that can reach superresolution \cite{Frustrated}. This subsequently led to the development of ``quantum imaging with undetected photons'' \cite{Zeiling,Machad,  Kviatovsky} and ``two-photon phase sensing with single-photon detection'' \cite{Vergyris2020b}. However, phase sensing beyond the classical limit was not demonstrated.

Here, we built upon these creative ideas and combine it with path-polarisation quantum state engineering \cite{Vergyris2020b} to show sub shot noise phase sensing, while pushing the sampling rates of previous experiments based on photon number states \cite{Cimini2019e,Ono2013c} by four orders of magnitude. Capitalising on these advances, we introduce a quantum optical microphone that surpasses the performance of an equivalent laser microphone. By benchmarking the microphone performance in a standardised audiology test on $n=45$ individuals, we provide a human experience of a quantum phenomenon, in other words, we make the quantum advantage audible.

The underlying idea of our sensing concept considers the canonical multi-photon path entangled state that acquires a phase: $|\Psi \rangle_{\text{path}} = (|N \rangle_\text{a}|\text{0} \rangle_\text{b} + e^{(i N \cdot \Delta \Phi)} |\text{0} \rangle_\text{a}|N \rangle_\text{b})/\sqrt{2} $, with $N$ being the photon number, while a and b denote the two different paths. This work focuses on the case $N=2$, for which we will show that, via quantum state engineering, we can transfer the two-photon phase information deterministically to a single-photon polarisation state: $|\Psi \rangle_{\text{path}} = (|\text{V} \rangle + e^{(i N \cdot \Delta \Phi)} |\text{H} \rangle)/\sqrt{2} $, with V and H denoting vertical and horizontal polarisation modes. At this point, the phase $N\cdot \Delta \Phi$ can be extracted from a simple intensity difference measurement after a polarising beam-splitter, in a one-to-one correspondence to classical interferometry based on single-photon interference. This has two critical implications: Basic intensity detection is sufficient such that no single photon or homodyne detection is required. Additionally, the difference intensity detection scheme can be efficiency used to reject common mode noise, thus providing improved robustness and reliability.
\FloatBarrier
\section{Phase measurement scheme}
\begin{figure}[htp]
	\includegraphics[width=0.45\textwidth]{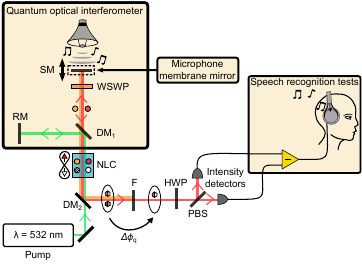}
	\caption{\label{fig_1} {\bf Advanced quantum optical sensor.}
		A 532 nm laser pumps a nonlinear crystal (NLC) to generate photon pair contributions in the forward and backward directions after reflection on the reference mirror (RM). The forward-generated pair contribution acquires the two-photon phase shift induced by displacements of the sample mirror (SM). The SM can be replaced by a microphone membrane mirror to implement an acousto-optical transducer for recording human speech. After a double passage through the wavelength-selective wave plate (WSWP), forward- and backward-generated pair contributions are overlapped without interference, which effectively transfer the two-photon phase shift onto a single-photon state. A filter (F) transmits only the relevant single-photon state from which two-photon phase shifts are reconstructed via simple light intensity detection at sampling rates up to $100\,\text{kHz}$. The difference intensity signal can be directly used to test the performance of the quantum microphone in a medical speech recognition test on humans. DM: Dichroic mirror splitting up pump laser light and photon pairs.
	}
	\end{figure}
The experimental setup is detailed in Fig. \ref{fig_1}. A 532 nm continuous-wave pump laser (green lines) is sent forwards through a periodically poled potassium titanyl phosphate nonlinear crystal (PPKTP), and is retroreflected at the reference mirror (RM) for a second backwards passage. Via type-0 collinear parametric down conversion in the PPKTP, each laser photon has a small chance to create a non-degenerate signal-idler photon pair contribution in either direction, $\rm |V,s \rangle_{\text{f}}|V,i \rangle_{\text{f}}$  or $\rm |V,s \rangle_{\text{b}}|V,i \rangle_{\text{b}}$, respectively (red and orange lines). The subscripts f and b indicate photon pair generation in the forward and backward direction, respectively, whose wavelengths are 1109 nm (s - signal) and 1023 nm (i – idler). By adjusting the pump power, we generate a photon pair flux of $\sim 1.65\cdot 10^8 \text{s}^{-1}$, which undercuts the parametric threshold value ($7.02\cdot 10^{12} \text{s}^{-1}$, see Supplementary Information), thus we operate in the regime of individual photon pairs.\\
The forward-generated photon pair contribution is retroreflected at the sample mirror (SM). Depending on the experiment, the SM is either a bulk mirror or a mirror membrane that is sensitive to sound waves. The distance at which the SM is placed with respect to the RM defines the two-photon phase shift $\Delta \Phi_\text{q} = \Delta \Phi_\text{s} +\Delta \Phi_\text{i}$. Here, $\Delta \Phi_\text{s}$  and $\Delta \Phi_\text{i}$  are the individual phase shifts accumulated by the signal and idler photon, respectively. One critical feature in our scheme is the dual-passage of the forward-generated pair contribution through a wavelength-selective wave plate (WSWP). The WSWP rotates, after a dual passage, the polarisation of the signal photon by $90^{\circ}$ while maintaining the idler photon polarisation, thus the quantum state is transformed to $e^{(i \Delta\Phi_\text{q})}|\rm \text{H},s \rangle_{\text{f}}|\text{V},i \rangle_{\text{f}}$. The advantage of the WSWP is that both photons in the forward-generated pair contribution accumulate the phase shift from the same path and are reflected at the same point on the sample mirror. Additionally, the creation of a cross-polarised state avoids interference when subsequently  overlapping forward- and backward-generated pair contributions. After overlapping both pair contributions into the same spatial mode, their subscripts f and b become obsolete, such that we have:
\begin{equation}
|\Psi \rangle_{\text{path}} =\left( \frac{|\text{V,s} \rangle + e^{(i\Delta\Phi_\text{q})}|\text{H,s}\rangle}{\sqrt{2}} \right)  |\text{V,i} \rangle.
\end{equation}
As intended, we have imprinted the two-photon phase shift $\Delta \Phi_\text{q}$ onto the signal single-photon state. As the idler photon does not carry useful information, we reject it using a $1109\pm 2\,\text{nm}$  bandpass filter. After the filter, phase information from the remaining signal photons is accessed via a projective measurement in the $\sigma_x$  basis. Experimentally, this is implemented using a half-wave plate (HWP), a polarising beam-splitter (PBS) and two intensity detectors $\text{D}_\pm$. The measured detector signals are:
\begin{equation}
	I_\pm \propto 1 \pm \nu_\text{q}\cos{(\Delta\Phi_\text{q})}
\end{equation}
in which the fringe visibility $\nu_\text{q}$  includes experimental imperfections, such as detector noise, loss, and non-optimal spatial mode overlap (see Supplementary Information).
The application-relevant key advantage of our scheme is that the two-photon phase $\Delta \Phi_\text{q}$  is now accessible with standard intensity measurements, including common mode noise rejection. Consequently, classical and quantum sensors can be directly compared using the same detection schemes, for example using detectors with high sampling rates.
\section{Benchmarking the quantum advantage}
As our quantum phase sensor resembles a Michelson-type interferometer, we naturally compare it to a classical Michelson interferometer operated with laser light at the average wavelength of the photon pairs ($1064\,\text{nm}$, for setup details, see Supplementary Information).
For both schemes, we define as a resource the amount of photons entering the interferometer, thus challenges associated with the quantum strategy are explicitly not corrected for (loss inside the interferometer and non-perfect mode-overlap inside the NLC).
Further, to benchmark the sensing schemes at the purest and fundamental level, it is important to minimise parasitic noise from the intensity detectors. This is why we use superconducting nanowire photon detectors (SNPD) for this demonstration. However, we emphasise that the SNPDs are operated as low-noise intensity detectors and that their single-photon detection capability is not required.
\begin{figure*}[htp]
	\includegraphics[width=0.9\textwidth]{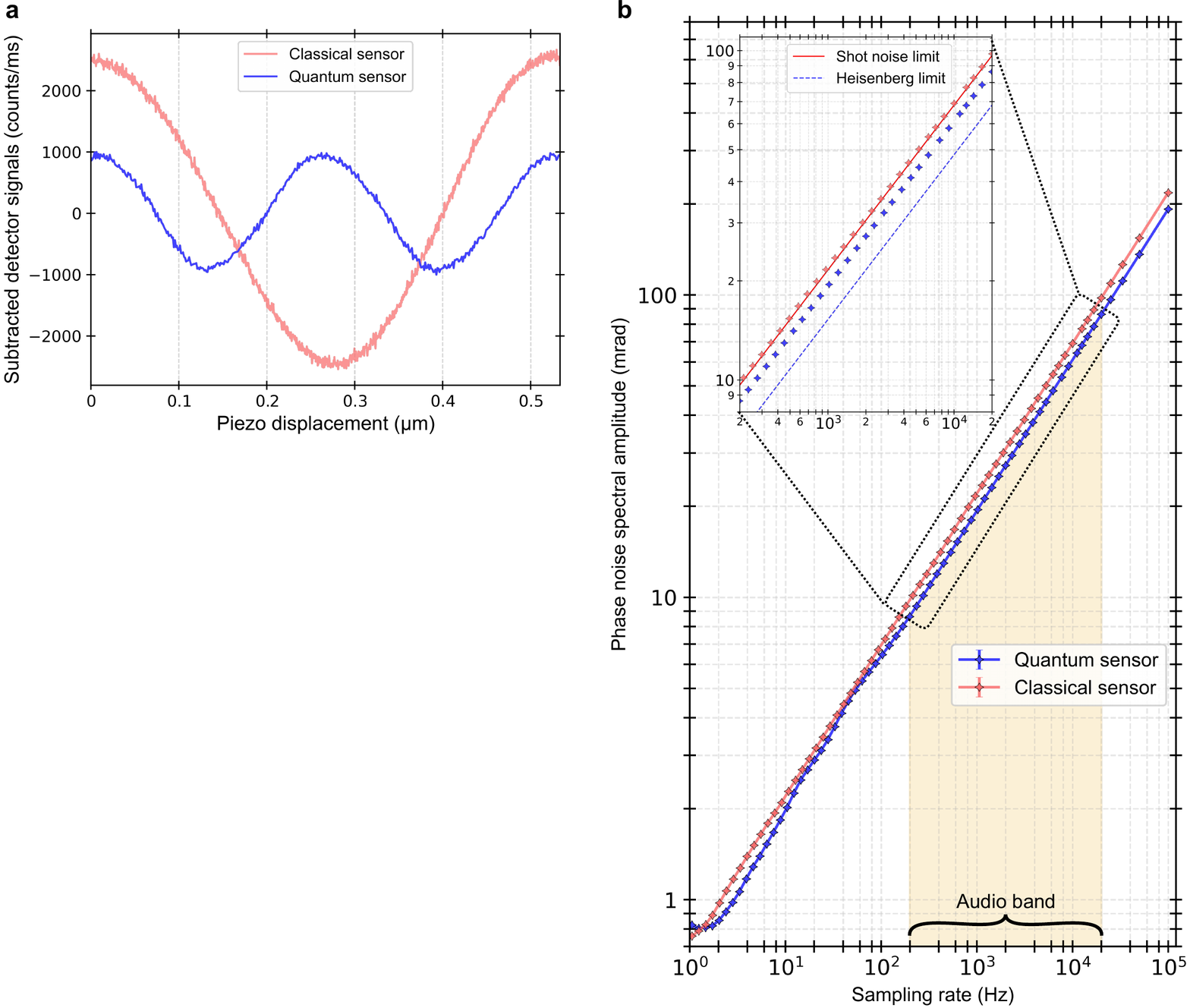}
	\caption{\label{fig_2a} {\bf Benchmarking the classical and quantum sensors.}
		{\bf a}, Subtracted detector intensity signals $D_+ - D_-$ for the classical (red) and the quantum sensors (blue). For the same piezo displacement, the quantum sensor presents twice as many fringes, thus corroborating phase sensing with super-resolution.
		{\bf b}, Frequency-dependent phase noise spectral amplitudes of the classical (red) and quantum (blue) sensor. Both measurements are performed at identical photon fluxes, prior to any loss inside the interferometer. Above $200\, \text{Hz}$, the quantum sensor surpasses the classical sensor by a factor of $1.13\pm0.02$. In the inset, the solid/dotted lines indicate the theoretical limits of both sensors at the used photon rate $R=2.14\cdot 10^6\, \text{s}^{-1}$.
		}
\end{figure*}

In a first step, we induce controlled phase-shifts in the classical and quantum interferometers by transducing the sample mirror with a piezoelectric actuator. The resulting difference intensities between both output detectors are shown in Fig. \ref{fig_2a}a. For the same piezo displacement, the quantum sensor displays twice as many interference fringes. This corroborates the use of path-entangled two-photon states that acquire phase information with super-resolution. However, super-resolution is not sufficient to demonstrate an exploitable quantum advantage \cite{Slussarenko2017c}. The critical parameter to consider is the achievable phase sensitivity per photon sent to the interferometer.

For the classical sensing scheme (subscript c), the optimal phase sensitivity $S_\text{c}$ occurs at the steepest slope in Fig. \ref{fig_2a}a, being $S_\text{c} = \sqrt{\eta_\text{ext}\eta_\text{int,c}\nu_\text{c}^2}^{-1}$. Here, $\nu_\text{c}$  is the fringe visibility, with $\eta_\text{int,c}$  denoting the mean optical transmissivity inside of the interferometer, and $\eta_\text{ext}$ is the external system efficiency outside of the interferometer (including optical filters, fibre-coupling and non-unity detector efficiency).
For the quantum scheme, the phase sensitivity at the steepest slope is $S_\text{q} = \sqrt{\eta_\text{ext}(\eta_\text{int,q}+1)\nu_\text{q}^2}^{-1}$ (the derivation is given in the Supplementary Information). Similarly to the classical sensor, $\eta_\text{int,q}$  specifies the single-photon transmissivity inside of the quantum interferometer. As both quantum and classical sensors share the same intensity detecting scheme, their external efficiencies $\eta_\text{ext}$  are the same. Consequently, the quantum sensor achieves an advantage over the ideal classical sensor ($\eta_\text{int,c}=1, \ \nu_\text{c}=1$), if
\begin{equation}
	\sqrt{(\eta_\text{int,q}+1)\nu_\text{q}^2}>1.
\end{equation}
Experimentally, we measure a raw fringe visibility of $\nu_\text{q} =0.85\pm 0.02$, mainly limited by imperfect photonic mode overlap and non-unity internal transmissivity $\eta_\text{int,q}$. In the worst case, the reduced visibility would be entirely determined by internal loss, which lower-bounds $\eta_\text{int,q} \geq 0.74$ (see Supplementary Information). This results in $\sqrt{(\eta_\text{int,q}+1)\nu_\text{q}^2}\geq 1.12>1$, thus demonstrating a quantum sensor advantage.

To comply with the demands of sensing applications, we experimentally infer the frequency-dependent phase sensitivities  $S_\text{c}$  and  $S_\text{q}$. To this end, we measure the spectral amplitudes of the phase noise for both the quantum and classical sensor. As shown in Fig. \ref{fig_2a}b, for sampling frequencies above $f=1\, \text{Hz}$, both sensors show the expected phase noise increase with a square root power scaling. The fit to the data shows that the classical sensor operates merely $1.5\%$ above the fundamental shot noise limit. Above $200\, \text{Hz}$, the quantum sensor surpasses the shot noise limit by a factor of $1.13\pm 0.02$, matching the predicted advantage of and  $S_\text{q} /S_\text{c}$. Here we note that the enhancement is sometimes defined as the squared ratio of signal-to-variance \cite{Hudelist2014a,Du2020b,G.FrascellaE.E.MikhailovN.TakanashiR.V.ZakharovO.V.Tikhonova2019,Fox2006}, but the associated enhancement factor $1.28\pm 0.04$ may be confusing as it does not directly reflect the phase sensitivity improvement.

A remarkable result is that our multi-photon quantum sensing scheme shows sub shot noise performance up to sampling rates of $f=100\, \text{kHz}$, which is at least four orders of magnitude higher than in previous approaches \cite{Ono2013c, Cimini2019e, Afek2}.

\section{A quantum optical microphone}
The much-increased frequency bandwidth of the quantum sensor widens significantly the range of possible applications. Capitalising on the quantum advantage across the entire audio-band of human speech (200 – 20000 Hz), we demonstrate the easy applicability of our sensor by implementing a quantum optical microphone, in analogy to a laser microphone \cite{Veligdan2000b}. Our intention is to make the quantum advantage audible, thus providing a human experience of a quantum phenomenon.\\
\begin{figure}[b!]
	\includegraphics[width=0.48\textwidth]{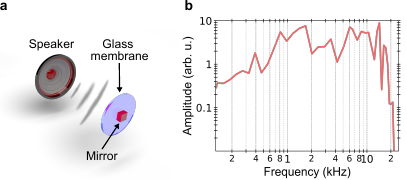}
	\caption{\label{fig_3} {\bf Acousto-optical transducer.}
		{\bf a}, A glass membrane with $12.7\, \text{mm}$ diameter and $70\, \rm \upmu m$ thickness translates acoustic pressure waves into a physical displacement. Photons are reflected by a small dielectric mirror that is glued on the membrane (mirror size $2\text{x}2\text{x}0.5\, \text{mm}^3$).
		{\bf b}, Frequency response of the transducer under single-tone excitation.
	}
\end{figure}

To convert acoustic waves to an optical signal, we develop an acousto-optical transducer. As shown in Fig. \ref{fig_3}a, we replace the SM in the interferometer by a $70\, \upmu\text{m}$ thin glass plate on which a small dielectric mirror is glued (mirror dimensions $2\text{x}2\text{x}0.5 \, \text{mm}^3$). We excite the membrane using a loudspeaker placed 3.5 cm behind the membrane. Single-tone frequency response measurements show a nearly flat response across the human audio band up to $15\, \text{kHz}$  (see Fig. \ref{fig_3}b). The membrane response linearity as a function of the speaker volume $V_\text{A}$ was additionally confirmed (see Supplementary Information).
We then use our setup to record audio snippets from a medically-approved speech recognition test (Oldenburger sentence recognition (OLSA) test \cite{Wagener1999c,Wagener1999f,Wagener1999g}, see Supplementary Information). The test is based on 600 words, distributed over 120 five-word long sentences. Words are played through the loudspeaker and recorded with the classical and quantum microphones at a sampling rate of $f=20\, \text{kHz}$. The signal-to-noise ratio of these recordings is adjusted in 22 steps via the loudspeaker volume $V_\text{A}$  measured in $\text{dB}_{\text{SPL}}$  (SPL = sound pressure level). In total, we record $2\text{x}22\text{x}600 =26400$ words. When expressing SNR and volume in decibels, we expect a linear relationship: 
\begin{equation}
	{\rm SNR}_i = \alpha_i V_\text{A}+\beta_i.
\end{equation}
Here, the subscript $i=\text{c},\text{q}$ denotes the classical and quantum microphone, $\alpha_i$ is the proportionality factor and $\beta_i$ is the SNR at $V_{\text{A}}=0\, \text{dB}_\text{SPL}$. Analysis of all audio data shows that both setups provide the same responsiveness: $\alpha_\text{c} = 0.95\pm 0.02, \ \alpha_\text{q}=0.95\pm 0.02$.  As expected, the quantum microphone’s baseline SNR is higher, i.e., $\beta_\text{c}=6.20\pm0.22\,\text{dB}_\text{SPL}$ and $\beta_\text{q}=7.04\pm0.20\, \text{dB}_\text{SPL}$, indicating already the quantum enhancement (see Supplementary Information).

To demonstrate that the quantum improvement is actually perceivable by humans, we subsequently conduct the medically-approved speech recognition test on $n=45$ subjects in a calibrated sound studio environment at the audiology department of the Olga hospital in Stuttgart (Germany). All individuals were German native speakers and normal hearing was initially verified by a pure tone audiometry test. To infer each subject’s individual speech recognition threshold (SRT, defined as the $50\%$ success threshold of understood words), they listen to randomised five-word sentences based on the German grammatical structure (name – verb – numeral – adjective – object). After communicating what they heard, the success percentage of understood words is noted. Repeating this procedure at different SNRs eventually results in a psychometric function from which one obtains the minimum required volume $V_\text{A}$ to reach the SRT. All subjects were familiarised with the test procedure by an initial training on ten sentences. Then, the SRT of each subject was inferred with $2\times 30$ sentences that have been recorded with the classical and quantum microphones. Although the OLSA test is resilient against training, we eliminate potential biasing by randomising the test order between subjects. Bias due to fatigue was avoided with a five-minute break after 30 sentences.\\
\begin{figure}[htp]
	\includegraphics[width=0.45\textwidth]{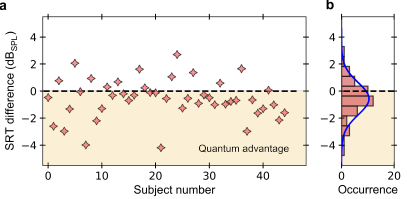}
	\caption{\label{fig_4} {\bf Results of the speech recognition test on $n=45$ individuals.}
		{\bf a}, Each individual’s difference in SRT for speech snippets recorded with the classical and quantum microphone, respectively. Data below the dashed line represent a quantum improvement.
		{\bf b}, Histogram of all data points. The quantum microphone reduces the volume level required to achieve the SRT by $-0.57\, \text{dB}_\text{SPL}$ with a $95\%$ confidence interval of  $\pm 0.44\, \text{dB}_\text{SPL}$ .
	}
\end{figure}

Fig. \ref{fig_4}a shows the difference of the SRT values between the quantum and classical microphones. Points below the difference SRT of $0\, \text{dB}_\text{SPL}$ indicate improved speech recognition performance with the quantum microphone. The data shows immediately that for the vast majority of subjects ($71\%$), the recordings with the quantum microphone led to a reduced threshold volume, in other words, a quantum advantage is observed. In Fig. \ref{fig_4}b, the data is grouped into a histogram, from which we deduce a mean improvement of $-0.57\, \text{dB}_\text{SPL}$, with a $95\%$ confidence interval of $\pm0.44\, \text{dB}_\text{SPL}$. The width of the histogram (one standard deviation) is $1.45 \, \text{dB}_\text{SPL}$, which is in accordance with the the systematic accuracy of single rounds of the OLSA test \cite{Wagener1999c,Wagener1999f,Wagener1999g}. Considering that we have performed $n=45$ measurements leads to an improvement of the relevance of our results, i.e., according to the square root law \cite{Freedman1976}, the quantum advantage is confirmed with a standard deviation of $(1.45 \, \text{dB}_\text{SPL} )/\sqrt{n}=0.22 \, \text{dB}_\text{SPL}$. Additionally, a statistical t-test with the hypothesis ``The quantum microphone does not result in an advantage'' is rejected with a $p$-value $p=0.006$, which is well-below the commonly accepted threshold value of $p=0.05$ (see Supplementary Information).
\FloatBarrier

\section{Conclusion}
We demonstrated a simple quantum-enhanced sub shot noise optical phase sensing scheme with common mode noise rejection at high sampling rates up to $f=100\, \text{kHz}$.
These results are enabled by path-polarisation state quantum engineering, which allows to measure signals with standard intensity detectors, thus overcoming the sampling rate issue in quantum optical sensing.
The possibility to use identical detection schemes in both classical and quantum sensing schemes makes it further straightforward to perform benchmarking. As such, we demonstrated a $1.13\pm0.02$ times sensitivity improvement for the quantum sensor and further improvements are realistic using optimised low-loss optics.

To highlight the reliability and applicability of the quantum sensing scheme, we then developed a quantum optical microphone for recording human speech. By comparing classical and quantum microphones at identical photon numbers, we showed that the quantum microphone records sound at a significantly reduced baseline noise level. These observations were subsequently confirmed in a medical speech recognition test on $n=45$ subjects, which showed that the speech recognition threshold was improved by $-0.57\, \text{dB}_\text{SPL}$ on average, thus confirming that humans could hear the quantum advantage. We believe that the combination of quantum sensors and human experiences is likely to trigger further interest in the field. 

As an outlook, we believe that our sensing scheme could be employed in applications involving biological samples, chemical reactions or atomic spin ensembles, which are sensitive to light exposure and/or short wavelength photons \cite{Wolfgramm2010a, Cimini2, fu, Li, Schermelleh, waldch}. In this sense, our quantum light source operates already within the favourable second biological window BW-II ($1000-1350\rm\,nm$) \cite{Rosal}. To show an exploitable quantum advantage, the light flux of our source would have to be increased to the level of a few 100\,nW. In that regard, a previous work demonstrated already photon pair beams at power levels of $0.3\, \upmu \text{W}$, while still operating in the district photon pair regime \cite{Dayan}.
At that point intensity detectors need to show a superior performance in opto-electronic conversion with noise figures of merit below the photonic shot noise. As we show experimentally in the Supplementary Information, suitable performance can actually be achieved by several commercially available InGaAs camera modules, thus putting quantum bio-imaging within realistic reach.

Additionally, we also believe that it would be also interesting to study the regime beyond distinct photon pairs fluxes and use the effect of stimulated emission similar as in SU(1,1) interferometry \cite{Chekhova2016c,Yurke1986,Ou2020f}. On one hand, higher power levels may further increase the quantum sensing enhancement factor, while on the other hand, path-polarisation state engineering may provide a pathway to introduce common mode noise suppression in these schemes.

Overall, we believe that our work provides a powerful asset to the toolbox of quantum optical sensing approaches, promising enhanced variety and advances in the field.

\FloatBarrier

\begin{acknowledgments}
The authors thank Barbara Baum, Annette Zechmeister, and Frank Schreiber for the participation in designing and manufacturing the mirror membrane used for this experiment. The authors also thank Louis Davide, Stefan Gaget and Markus Kleinhansl for helpful discussions regarding the statistical evaluation, Marwa Garsi and Katharina V. Wutz for counselling regarding the design of the figures and Élie Gouzien for fruitful discussions regarding the theoretical framework. We thank Andreas Vollmer for support on the measurements on the WSWP performance. We thank Lukas Niechziol for proofreading and fruitful discussions. Daniil M. Lukin acknowledges the U.S. Department of Energy, the Office of Science, under Awards DE-SC0019174 and DE-Ac02-76SF00515. Stefanie Barz acknowledges support from the Carl Zeiss Foundation, the Centre for Integrated Quantum Science and Technology (IQST), the German Research Foundation (DFG), the Federal Ministry of Education and Research (BMBF, project SiSiQ), and the Federal Ministry for Economic Affairs and Energy (BMWi, project PlanQK). Florian Kaiser, J\"org Wrachtrup and Raphael Nold acknowledge the support of the Centre for Integrated Quantum Science and Technology (IQST), the GRK 2642, the European Science Council via ERC grant SMeL and the Max Planck Society.
\end{acknowledgments}

\section*{Author Contributions}
R.N., J.S. and T.L. performed the experiment; R.N. and F.K. analysed the data. F.K. designed the experiment and provided experimental assistance. R.N. and F.K. wrote the paper. F.K., J.W. and C.M.G. supervised the project. R.S., D.M.L., V.V., C.B. and J.V. assisted with discussions and design ideas to the development of the mirror membrane. M.T.P.Z. and R.B. provided helpful discussions about the statistical data analysis. C.B. provided the theoretical framework of the experiment. S.B. provided laboratory space for recording the sound files. R.B. provided the sound studio environment to carry out the OLSA tests. R.K. and S.B. developed and provided experimental equipment. R.K. and R.N. built up the classical laser source. All authors provided helpful comments on the manuscript.

\section*{Correspondence and requests for materials}
should be addressed to f.kaiser@pi3.uni-stuttgart.de.

\newpage

\newpage

\end{document}